\documentclass[onecolumn,preprint,10pt]{elsarticle}

\usepackage{amssymb}
\usepackage{graphicx}% Include figure files
\usepackage{dcolumn}% Align table columns on decimal point
\usepackage{bm}% bold math
\usepackage[section]{placeins}% for using \FloatBarrier
\usepackage{array} % for the table
\usepackage{subfigure}
\usepackage{pxfonts} % for upper case roman latters
\usepackage{multirow}
\usepackage{array}

\journal{Physics Letters A}

\begin{document}

\begin{frontmatter}

%% Title, authors and addresses

%% use the tnoteref command within \title for footnotes;
%% use the tnotetext command for the associated footnote;
%% use the fnref command within \author or \address for footnotes;
%% use the fntext command for the associated footnote;
%% use the corref command within \author for corresponding author footnotes;
%% use the cortext command for the associated footnote;
%% use the ead command for the email address,
%% and the form \ead[url] for the home page:
%%
% \title{Title\tnoteref{label1}}
% \tnotetext[label1]{}
\author{Kun Fang}
\author{G. W. Fernando}
% \ead[url]{home page}
\address{Department of Physics, University of Connecticut, Storrs, CT 06269, USA}
\author{A. N. Kocharian\corref{cor}}
\address{ Department of Physics, California State University, Los Angeles, CA 90032, USA}
\cortext[cor]{Corresponding author at: Department of Physics, California State University, Los Angeles, CA 90032, USA}
\ead{armen.kocharian@calstatela.edu}
% \fntext[label3]{}

\title{Pairing enhancement in  Betts
%remove "finite"
finite
%square
lattices with next nearest neighbor couplings: exact results
}

\begin{abstract}
Electron instabilities in the Hubbard model with the next nearest neighbor coupling are calculated by exact diagonalization in finite, two-dimensional Betts cells (lattices). A viable {\sl spin } and {\sl charge} coherent pairing, signaled by quantum critical points and the negative charge gap region, is found in
 8- and 10-site Betts lattices at small and moderate $U$ regions consistent with our exact results in elementary bipartite geometries [Phys.~Rev. B{\bf 78}, 075431 (2008)]. The contour isolines for continuous temperature driven-crossover between the Mott-Hubbard insulating and coherent pairing phases are demonstrated. The criteria for smooth and abrupt phase transitions are found for systematic enhancement of coherent pairing by optimization of the next nearest neighbor coupling parameter.
\end{abstract}

\begin{keyword}
high $T_c$ superconductivity \sep charge pairing \sep spin pairing, Betts lattice
%% keywords here, in the form: keyword \sep keyword
%% MSC codes here, in the form: \MSC code \sep code
%% or \MSC[2008] code \sep code (2000 is the default)

\end{keyword}

\end{frontmatter}
\newcommand{\rar}{\rightarrow}
\newcommand{\ket}[1]{\left| #1 \right\rangle}
\newcommand{\bra}[1]{ \left\langle #1 \right|}
\newcommand{\la}{\left\langle}
\newcommand{\ra}{\right\rangle}

%\nofiles

\maketitle

\section{Introduction}~\label{intro}
A key element for understanding the physics in cuprates, pnictides, manganites and colossal magnetoresistant (CMR) nanomaterials is the experimental observation of magnetic and density phase separation instabilities at the nanoscale~\cite{Anderson,Tranquada,Dagotto}. The spatial inhomogeneities with short range electron correlations in the absence of long-range order play a crucial role in determining the mechanism of pairing instabilities in the high $T_c$ superconductors (HTSCs) and CMRs. Therefore, the local nature of the (pairing) interaction is sufficient for describing electron pairing instabilities
%within
in the framework of the
two-dimensional (2d) Hubbard model.
 The finite-cell-based lattice or small-size optimized clusters may be one of the few {\it solid grounds} available to handle this challenging issue~\cite{JMMM,PRB,Kivelson}. A new, emerging guiding principle for the search of new materials can be identified as  spatial inhomogeneities and {\it density} phase separation instabilities in the proximity to quantum critical points (QCPs). Strong quantum fluctuations can even dominate thermal fluctuations and affect the properties of a material well above absolute zero~\cite{Fisk} (see Ref.~\cite{Springer} and also references therein). The phase separation instabilities in small (bipartite) Hubbard clusters studied in Refs.~\cite{AKA08P,GW07,PLA07} display QCPs and interesting thermodynamics, which depend on the strength of the on-site Coulomb repulsion ($U$), the cluster topology and temperature. In contrast, spontaneous transitions in frustrated (non-bipartite) geometries can occur for all $U$ by avoiding QCPs (level crossings) at finite $U$ and such transitions depend strongly on the sign of the hopping term $t$.

The existence of the QCPs associated the phase separation instabilities may be a crucial ingredient of the superconducting transition and can provide important clues for understanding the {\it incipient microscopic} mechanisms of charge and spin pairing instabilities in the HTSCs and CMRs. Here our primary focus is the mechanism of electron pairing due to the density instabilities in small bipartite lattices in the ground state and finite temperatures, where there is no general agreement on the phase separation boundaries; related results are still controversial, especially at small $U$ values~\cite{Dagotto2,Emery,Hellberg}. Small systems suffer from finite-size  and edge effects, so it is unclear whether the observed instabilities can survive in the thermodynamic limit. Thus, tests on reduced boundary effects are necessary to confirm the picture of instabilities in small cluster-based larger systems in the so called "optimized" Betts lattices~\cite{betts}.

\begin{figure}
\includegraphics*[width=20pc]{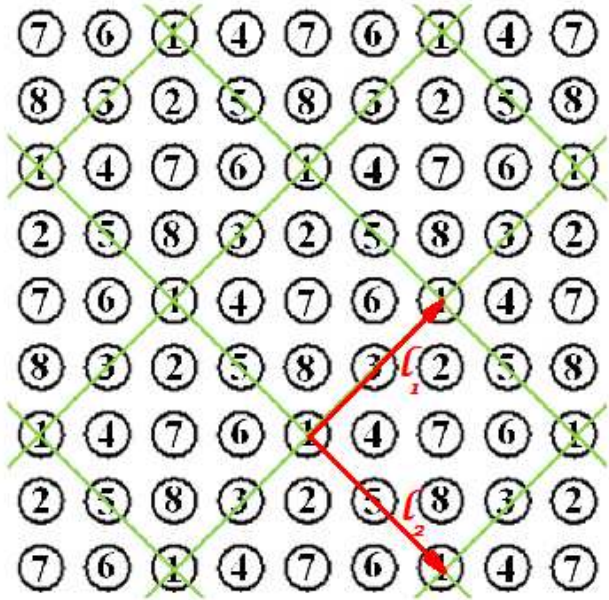}
\caption{The 8-site finite unit cell ({\it plaquette}) for the square lattice. When repeated periodically, it can fill the entire ({\it infinite}) space. The cells have ${\bf l}_1$ and ${\bf l}_2$ edge vectors, (2,2) and (2,-2), as defined in Ref.~\cite{betts96}.}
\label{fig:lattice}
\end{figure}

The exact ground state properties of the infinite two-dimensional (2d) Heisenberg-like lattice have been extracted from systematic studies of finite Betts clusters~\cite{betts96}. An infinite square lattice can be tiled by squares of $L$-site clusters with edge vectors~\cite{betts} which represent displacements of one vertex to the equivalent vertex in the neighboring tile with which it shares an edge (See Figure ~\ref{fig:lattice}). Different block structures of ``square symmetry" broken up into nearly decoupled clusters can be used as plaquettes to extrapolate the results to the infinite square lattice. Studies of finite clusters with periodic boundaries and next nearest neighbor hopping can play an important part in testing the reliability of the results drawn from finite-size systems\cite{Oitmaa,Hirsch}. The phase separation boundaries have been calculated within the $t$-$J$ model~\cite{Kaxiras,Kaxiras1}, although the validity of the strong-coupling expansion has been questioned~\cite{Friedman}. As far as the authors are aware, the exact studies of pairing thermodynamics have not been attempted even in small Betts-cluster-based lattices neither in the framework of single orbital Hubbard model with or without the next nearest neighbor coupling.

In this paper, we address two specific issues: (i) check how exact calculations in finite Betts-cluster-based lattice are consistent with our exact results in 2${\times}$2 and 2${\times}$4 clusters and, (ii) analyze the effect of lattice frustration on phase separation instabilities in bipartite geometry driven by the next nearest neighbor coupling. A related question is whether the physical properties extracted from these Betts square lattices are likely to be applicable in the infinite lattice~\cite{betts96}. In Sec.~\ref{meth}, the square geometry of the Betts lattice is explained and the corresponding Hamiltonian is described. In Sec.~\ref{result}, the exact results on charge and spin pairing in the Betts lattice based on canonical and grand canonical ensemble are analyzed and corresponding key physical properties are calculated. The finite temperature effects near QCPs at various $U$ values are discussed in Sec.~\ref{finite_temp}. The enhancement of pairing driven by the next nearest neighbor coupling is studied in Sec~\ref{nnn}. The conclusions in Sec.~\ref{conclude} provide a summary of what was learned from exact studies of electron pairing instabilities in the presence and absence of electron-hole symmetry.

\section{Betts square lattice}~\label{meth}
As shown in Figure~\ref{fig:lattice} the entire 2d lattice space can be tiled by periodically repeated small Betts squares, which are connected through edge vectors ${\bf l}_1$ and ${\bf l}_2$. Such periodic boundary conditions reduce the edge effects while taking full advantage of the space group symmetries of the isotropic two dimensional (square) lattice. Notice that, in the 8-site Betts lattice, each odd site is surrounded by all the even sites (as nearest neighbors) and vice versa. This structure illustrates the bipartite character of Betts (square) cells. For comparison, we study the effect of frustration on electron pairing in the absence of electron-hole symmetry by introducing the next nearest neighbor coupling $t_{nnn}$ with periodic boundary conditions. The model describes the electron doping for $t_{nnn}>0$ and the hole doping regime for $t_{nnn}<0$.

\subsection{Model}
We consider the extended Hubbard model
\begin{eqnarray}
H=-\sum\limits_{ i,j \ \sigma}t_{ij} c^{+}_{i\sigma}~c_{j\sigma}+U \sum\limits_{i} n_{i\uparrow}n_{i\downarrow}
\label{eqn:h}
\end{eqnarray}
where summation over $i$ and $j$ in Eq. (\ref{eqn:h}) goes through all lattice sites $L$ with coupling integral $t_{ij}$
\begin{eqnarray}
t_{ij} =\left\{ {\matrix{\displaystyle{t}\cr \displaystyle{t_{nnn}}\cr \displaystyle{0}\cr} \ \matrix{{\ \ \ \ \ \
\ {\rm if } \ \ \ {\rm i }, {\rm j }\ \ {\rm are }\ \ {\rm nearest }\ \ {\rm neighbors }},\cr {\ \ {\rm
if}\ \ {\rm i }, {\rm j }\ \ {\rm are }\ \ {\rm next}{\rm -nearest }\ \ {\rm neighbors }},\cr {\  {\rm
otherwise,}\ }\cr} } \right. \label{sU0}
\end{eqnarray}
and $U>0$ is the on-site Coulomb interaction. The energies are measured with respect to $t_{nn}>0$, which is set to 1 everywhere, unless otherwise stated.

\begin{figure}
\includegraphics*[width=20pc]{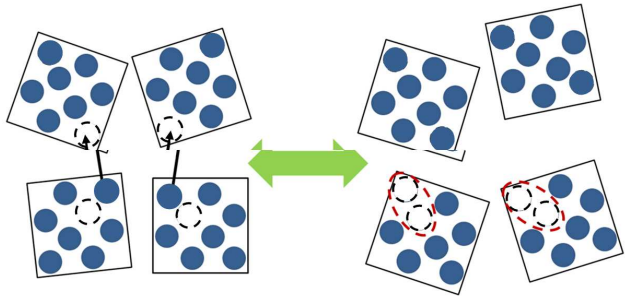}
\caption{Schematic drawing of hole (electron) redistribution within Hubbard nanoclusters with one hole of half-filling in the grand canonical ensemble of 8-site clusters at low temperatures. The state on the left is an ordinary state at $N=7$ per unit cell. Notice that the spontaneous fluctuations in particle number, near the average $\left\langle N\right\rangle=7$, are energetically  favorable and make electron redistribution across the ensemble of clusters possible even without direct contact (hopping) between them. When $\Delta^c>0$, the hole localization on separate clusters corresponds to a
homogeneous stable Mott-Hubbard-like insulating $d^7$ state with electron-hole pairing. When $\Delta^c<0$, the hole pair binding in $d^6$ and half-filled $d^8$ inhomogeneous cluster configurations on the right are energetically preferred.}
\label{fig:pair}
\end{figure}

\section{Results}~\label{result}
\subsection{Phase separation instabilities}~\label{nn}
We consider $L$=8 and 10 site Betts clusters where the dimensions of the Hilbert spaces are $4^8$ and $4^{10}$ respectively. An exact diagonalization along with the Lanczos algorithm~\cite{Lanczos} is employed to evaluate the relevant, low lying eigenstates at all filling (doping) values. These eigenstates are used in Sec.~\ref{charge} and ~\ref{spin} to extract the charge and spin pairing gaps and corresponding pairing instabilities of the model.

\subsubsection{Charge pairing instability}~\label{charge}
The charge pairing instability in the negative charge gap region, as defined below, is a precursor to a spin pairing instability at rather low temperatures (see Sec.~\ref{spin}). In our aforementioned publications, we have discussed this issue in selected cluster geometries such as the $2\times2$ square and  $2\times4$ ladder. We identify charge pairing behavior at zero and finite temperatures in a $L$-site cluster by defining a charge gap $\Delta^c(N,T)$ at a given $U$ in a particular doping region:
\begin{eqnarray}
\Delta^c(N,T)=E(N+1,T)+E(N-1,T)-2E(N,T)
\label{eqn:cg}
\end{eqnarray}
where $E(N,T)$ is the lowest canonical many body energy for an $N$-electron state at a fixed temperature $T$. This charge excitation gap determines the stability of an $N-$electron state compared to an equal admixture of $(N+1)-$ and $(N-1)-$electron states. As shown  in Figure ~\ref{fig:pair}, the electron number in clusters  placed in a thermal reservoir can fluctuate near an average number $\left\langle N\right\rangle$. This schematic picture for $\left\langle N\right\rangle=7$ illustrates the {\it incipient} mechanism of hole pairing at the 8-site cluster level, where properties such as density can strongly fluctuate. (These fluctuations closely resemble interconfiguration fluctuations in mixed valence compounds~\cite{koch}.) Eq. (\ref{eqn:cg}) describes an interconfiguration energy gap for electron fluctuations between different
 many body states and cluster configurations ($d$) that differ in electron number $N$~\cite{PLA09},
\begin{eqnarray}
d^N +d^N=d^{N+1}+d^{N-1}.
\label{eqn:config}
\end{eqnarray}

\begin{figure}
\includegraphics*[width=20pc]{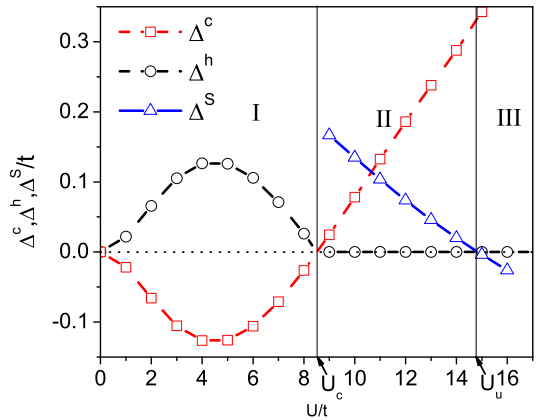}
\caption{The {\it canonical} charge $\Delta^c$ and spin $\Delta^s$ excitation gaps in the ensemble of the 8-site Betts cluster for $N=7$ as a function of $U$ at $t_{nnn}=0$ and infinitesimal $T\to 0$. The {\it grand canonical} spin gap $\Delta^h>0$ provides the stability for spin singlet pairing at $0<U<U_c$. The nodes $\Delta^{c,s,h}(U)=0$ of the charge and spin gaps define a quantum critical points for various electron instabilities. This phase diagram displays the main phases found earlier for elementary square geometry in Ref.~\cite{AKA08P}. Phase I is a phase with a negative charge gap and a positive spin gap of equal amplitude ($\Delta^c=-\Delta^h$) at $U<U_c$
 which  describes  coherent pairing of electrons (holes) in the spin singlet state. Phase II is a spin liquid phase with gapless, low spin-${3\over 2}$ excitations separated from the higher spin-${5\over 2}$ by canonical gap $\Delta^s>0$. In Phase III, the negative spin gap manifests the onset of low spin-${5\over 2}$ (unsaturated) ferromagnetism. ($U_u$ identifies the crossover point for the spin-z 3/2 to 5/2 transition.) There is also a consequent transition into a fully saturated spin-${7\over 2}$ ferromagnetism at larger a $U$ value, $U_s$ (not shown).}
\label{fig:gapT0}
\end{figure}

Depending on the strength of the on-site electron-electron repulsion $U$, this charge excitation gap can be positive $\Delta^c>0$ or negative $\Delta^c<0$. Physically, $\Delta^c>0$ manifests a stable $d^N$ (cluster) configuration, while $\Delta^c<0$ describes a first order phase separation instability with spontaneous generation of $d^{N-1}$ and $d^{N+1}$ electron (cluster) configurations. Thus, a negative charge gap ($\Delta^c<0$) implies an effective attraction or pairing between holes accompanied by a (spontaneous) transition into an inhomogeneous phase, while a positive $\Delta^c$ shows a preference for holes to be localized in separate clusters which are intimately related to insulating behavior and the Mott-Hubbard-like electron-hole pairing~\cite{PLA09}. For a given chemical potential and temperature in the grand canonical ensemble, electrons are allowed to be redistributed among the clusters to optimize their free energy. Thus, on the left of Figure ~\ref{fig:pair}, a homogeneous Mott-Hubbard state with electron-hole pairing is preferred in the positive charge gap ($\Delta^c>0$) region, while in the negative gap region ($\Delta^c<0$) on the right, an inhomogeneous phase with hole-rich ($N=6$) and hole-poor ($N=8$ half-filled) clusters becomes energetically favorable.

\begin{figure}
\includegraphics*[width=20pc]{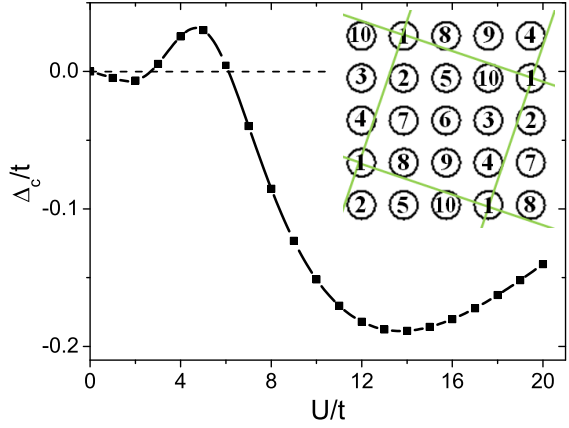}
\caption{A plot of the charge gap $\Delta^c$ in the 10-site Betts cell at $N=9$ as a function of $U$ at $t_{nnn}=0$ and $T=0$. In contrast to the 8-site lattice, optimized pairing is shifted to larger $U$ values and a positive charge gap appears at very small $U$ values. This is due to the
reduced
symmetry of 10-site geometry (see the inset) compared with the 8-site lattice (see Sec.~\ref{symmetry}).}
\label{fig:gap_10}
\end{figure}

The charge gap $\Delta^c$ at $T=0$ is plotted as a function of $U$ in Figure~\ref{fig:gapT0} for $N=7$. The nodes for the sign change in the excitation gap provide strong evidence for the existence of QCPs for a level crossing instability, associated with a first order phase separation transition~\cite{AKA08P}. For example, the charge gap in Figure ~\ref{fig:gapT0} vanishes ($\Delta^c (U_c)=0$) at $U=0$ and also at a particular $U$ value, {\it i.e.,} $U_c=8.52$. These QCPs describe an abrupt (discontinuous)
 change in the ground state of a many-body system due to strong quantum fluctuations. Compared to the elementary square geometries, the negative charge gap ($\Delta^c<0$) for the 8-site clusters with electron number $N=7$ has a larger
(antinode gap) 
maximum binding energy and shows a larger region $0<U<U_c$ of electron instability. The Mott-Hubbard insulating ground state for electron-hole pairing with $\Delta^c>0$ is stable at all $U>U_c$ (shown in Fig.~\ref{fig:gapT0}) while for $0<U<U_c$, the ground state consists of paired electrons. This instability in redistribution of electrons, caused by electron hopping between clusters, signals the formation of inhomogeneous, hole-poor $d^{N-1}$ and hole-rich $d^{N+1}$ regions or indicates a tendency toward phase separation~\cite{PLA09}.

\subsubsection{Square symmetry}~\label{symmetry}

Broken symmetry of the two-dimensional planar geometry in various finite-size
Hubbard
clusters plays a crucial role in determining the pairing symmetry and superconducting properties. Our results in various square (bipartite) geometries show that pairing properties strongly depend on the cluster symmetry. The gap behavior in the 8-site Betts
%clusters
cells is different from the 2${\times}$4 ladder~\cite{GW07,PLA07}. For comparison, in Figure ~\ref{fig:gap_10}, we show the exact charge gap with reduced square symmetry in
 the 10-site Betts lattice as a function of $U$. The shallow charge gap at small $U$ is transformed to a significantly 
 larger one with maximum (antinode) binding gap energy
 at moderate $U$. This figure also shows an oscillating gap due to the reduced square symmetry which resembles that of the $2\times4$ clusters~\cite{GW07}. Both the $2\times4$ ladder and the 10-site Betts lattice
(with edge vectors (1,3) and (3,-1))  have lower symmetries than the 8-site Betts
%lattice
cell
(with edge vectors (2,2) and (2,-2))
 and we believe that the oscillations in the charge gap (as a function of $U$) are related to this. Although a positive charge gap appears at very small $U$ values,
 in overall,
 the evolution of the gap is quite similar to our results obtained in other clusters, except at very low $U$.
Thus, Betts 8- and 10-site clusters provide strong support for electron instabilities and nanoscale inhomogeneities found in generic $2\times2$ and $2\times4$ clusters in ~\cite{GW07,PLA07} and reproduced later~\cite{Kivelson2}.

\subsubsection{Spin (pseudo)gaps}~\label{spin}
%Spin gaps
%%%%%% new section -- completely replaces the previous sections on spin gaps
% Armen
Spin (pseudo)gaps classified according to their total spin $S$ give essential insights into the
behavior of the system
on response to applied magnetic field $h$
in the vicinity of one hole off half filling
%(say
%$\mu=\mu_c$)
%
%in regions where the charge gap is
for negative ($U<U_c$)
%(for the 8-site Betts lattice, $0<U<U_c$)
or positive ($U>U_c$)
%($U>U_c$ for the 8-site Betts lattice)
charge gaps.
%This is simply related to having a ground state with total spin zero or nonzero respectively.
This is simply related to a ground state having total spin
either
zero or nonzero respectively.

Using exact analytical expressions for {\it the grand canonical potential} we analyze the variation of the spin $S^{z}(\mu)$ and susceptibility $\chi_h={{\frac {\partial  S^{z}(\mu)} {\partial h}}}$ as a function of magnetic field $h$ near the critical chemical potential, $\mu_c$, close to given average electron number, $\left\langle N\right\rangle=L-1$.
In the negative charge gap region, near $\mu_c$, the ground state has zero total spin, $S=0$. In order to break spin pairs and generate spin excitations, a finite magnetic field need to be applied. The threshold value of this field will be identified as a
%pseudo spin gap
spin pseudogap
%($\Delta_h$) as described in fig 3
($\Delta^h$), as described in our previous publications~\cite{AKA08P, GW07}. This spin
% gap
pseudogap
at $0<U<U_c$ provides (rigidity) stability
%($\Delta_h>0$) as defined on fig 3
($\Delta^h>0$)
to paired (singlet) spins in the ground state. At rather low temperatures, near
% $T=0$
$T\to 0$,
the stable spin pseudogap is equal in magnitude to the charge pairing gap, $\Delta^P=-\Delta^c$. This is an important characteristic of phase coherence in the ground state associated with simultaneous coherent pairing of independent charge and spin (Fermion) entities and their corresponding full Bose-Einstein condensation (real-space pairing) with a single energy gap. Such behavior is similar to the coherent pairing with unique
quasiparticle gap
in conventional BCS theory~\cite{AKA08P,PLA09}.
%However, it
It is important to note that the coherent pair formation through {\sl separate} condensation of bound holes and coupled opposite spins
%(bosons)
(both bosons)
at different critical temperatures ${T_c}^P$ and ${T_s}^P$ in real space here is somewhat different from the Cooper %pairing
pairs (composite bosons) condensation in momentum space with
%a single quasiparticle gap and
the unique critical temperature.
%, T$_c$.
%In addition, the
The magnitudes of the spin pseudogap and charge gap
also
differ with increasing temperature here, unlike in the BCS case. It is consistent with ARPES data that excitations in high $T_c$ superconductors (as in the exact solution) are not well defined quasiparticles  as
it is
in the BCS theory~\cite{Springer}.

In the positive charge gap $U$ region(s), the behavior of states near $\mu_c$ is different from the above due to the
presence of
a spin-charge separation effect~\cite{AKA08P} in the
ground state that carries a nonzero total spin, $S\neq 0$. Clearly, this ground state can be excited with an infinitesimal magnetic field, hence spin excitations here are gapless. In addition to the pseudospin gap $\Delta_h$, one can also focus on the total spin
%S
$S$
of the ground state which has a spin degeneracy of
%(2S+1)
(2$S$+1). It turns out that, depending on the value of
%U,
$U$
due to level crossings, certain
competing
states with high or low (total) spin can become the ground state. Hence, we define a {\it canonical} spin gap $\Delta^s$ as the energy difference between the ground state with spin $S$ and the lowest excited state with spin $S^{'}$ having the same $N$ and $U$ as
\begin{eqnarray}
\Delta^s(N,T)=\left\{ {\matrix{\displaystyle{ E(N,S^{'},T)-E(N,S,T)}\cr
 \displaystyle{E(N,S,T)-E(N,S^{'},T)}\cr} \ \matrix{{\ \ \ \
  {\rm when}~~ S^{'}\ge S},\cr {\ \
~~ {\rm when}~~ S^{'} < S},\cr} } \right.
\label{eqn:sg}
\end{eqnarray}
where $E(N,S,T)$ is the canonical energy of an $N$ electron state with  total spin $S$ at rather low temperature $T$. If the excited state has a higher (lower) total spin, the spin gap is positive (negative). This gap (along with its sign) identifies whether a high or low spin state turns out to be the lowest average energy state (or ground state at
%$T=0$
$T\to 0$). In addition, the ground state here can be described as a spin liquid due to the
spin
degeneracy
%(2S+1).
(2$S$+1), which can be lifted by infinitesimal $h\to 0$.
One other noteworthy point is that, in the 8-site Betts lattice, there are several
%U
$U$
values (QCPs) at which such level crossings occur (from low-spin to high-spin or vice versa) near $\mu_c$. Those are denoted by $U_c$ (spin 0 to 3/2), $U_u$ (from spin 3/2 to 5/2) (see Fig.~\ref{fig:gapT0}) and $U_s$ (from spin 5/2 to 7/2).

\subsubsection{Finite temperature effects}~\label{finite_temp}
\begin{figure}
\begin{center}
\end{center}
\includegraphics*[width=20pc]{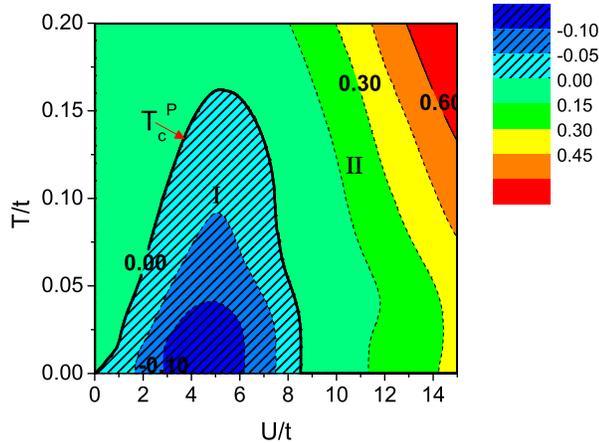}
\caption{A contour plot of charge gap in the 8-site (periodic) Betts cluster at $N=7$ as a function of Coulomb repulsion $U$ and temperature $T$. As $T$ increases, the region of negative charge gaps becomes smaller and the crossover point $U_c(T)$ shifts to lower $U$ values. The negative gap disappears completely at about $T=0.16$. The solid boundary in the figure denotes the contour $T^P_c(U)$ at which the charge gap vanishes ($\Delta^c=0$). Phase I with a shaded pattern is the electron-electron pairing phase. Phase II without a shaded pattern is the electron-hole pairing phase.}
\label{fig:cgapTn0}
\end{figure}

The QCPs introduced in Secs.~\ref{charge} and ~\ref{spin} turn out to be useful for the analysis of spontaneous instabilities at nonzero temperatures due to the interplay of quantum and thermal fluctuations in proximity to the above discussed QCPs. These QCPs as well as the doping dependencies on the chemical potential at nonzero crossover temperatures (not shown here) are qualitatively similar to those obtained earlier~\cite{PLA09}. A contour line (isoline) for the charge gap as a function of $T$ and $U$, along which the gap attains a constant value $c$ ($\Delta^c(U,T)=c$) defines the contour map
in
Figure~\ref{fig:cgapTn0}. The
slice
cut of the vanishing gap ($c=0$) at finite temperatures ($\Delta^c(U,T)=0$) defines the boundary between positive and negative charge gap phases, which is marked by the solid line in  Figure ~\ref{fig:cgapTn0}. This crossover is a smooth second order transition for the onset of pairing instabilities at finite temperatures. For a given $T$, the 
%width 
magnitude and area
of the negative gap region are an indication of the pairing strength and the region of the phase separation instability; this
%width 
magnitude and area
shrink and disappear at relatively low temperature, indicating that the pairing and phase separation instability 
is a low temperature phenomenon. 
 The solid boundary line evolves smoothly with temperature and approaches the QCPs, $U=0$ and $U_c=8.52$, as temperature goes to zero, $T\to 0$. Exactly at $T=0$, this picture shows the region $0\leq U \leq U_c$ for coherent pairing. The negative gap region defines first order phase separation instabilities, while the positive gap is a signature for smooth second order transitions. Notice that the electron pairing gap below ${T_c}^P(U)$ with $\Delta^c(T)<0$ is transformed into a Mott-Hubbard gap $\Delta^c(T)>0$ when the temperature increases above ${T_c}^P(U)$. This picture describes the transition from incoherent charge pairing of preformed electron (hole) pairs with decoupled spin into a Mott-Hubbard insulator. Thus, the continuous transition at the classical critical temperatures here from one, ordered, low-temperature phase into another (quantum) ordered high temperature phase is somewhat different from the conventional phase boundary described by an order parameter, which is non-zero in the ordered phase and zero in the disordered high temperature phase. Note, that the high temperature insulating phase in Figure~\ref{fig:cgapTn0} at ${T_c}^P(U)$ in $0<U<U_c$ region originates from the low temperature (quantum) Mott-Hubbard phase at $U>U_c$. Thus, we conclude that the QCP here is unique, since it describes a boundary for a smooth transition between two quantum ordered phases with different symmetries. The effect of quantum criticality in Figure~\ref{fig:cgapTn0} can be felt at low temperatures without even ever reaching the ground state ($T=0$).

\subsection{Next nearest neighbor coupling}~\label{nnn}
In many real materials (such as the cuprates), the contribution of hopping among next nearest neighbor (nnn) atoms can be important due to many reasons, such as its ability to separate hole and electron doped properties by breaking particle-hole symmetry ~\cite{Duffy}. Experimental results strongly support the idea that holes in CuO$_2$ planes are mainly sited in the $d_{x^2-y^2}$ Cu orbital in related undoped insulating materials, and additional holes are transferred mainly to the oxygen p orbitals upon doping. The next-nearest coupling term has already been invoked to explain the shapes of the Fermi surfaces obtained from recent band-structure calculations for these new superconducting materials. The two orbital Hubbard model with  p-bonding orbitals, after elimination of the Cu sites, can be reduced to a single-orbital Hubbard model 
with nearest and next-nearest-neighbor couplings 
(\ref{sU0})
by allowing holes to move within a given oxygen sublattice. The pairing can be changed significantly by the Coulomb repulsion, when the second-neighbor interactions are included. Therefore, a study 
%with next nearest neighbor hopping 
in the Hubbard model (\ref{sU0}) with $t_{nnn}\neq 0$ might give a more realistic physical picture related to real materials. In this section, we discuss the effects due to a nonzero $t_{nnn}$. Because
%the nnn hopping 
$t_{nnn}$
in real material is usually much smaller than $t$ parameter 
%the nearest neighbor hopping, 
we only show results for 
%relatively small nnn hopping parameters ($-0.3<t_{nnn}<0.3$)
$-0.3<t_{nnn}<0.3$. Below we will find conditions under which broken $C_4$ symmetry in
 frustrated Betts cells with $t_{nnn}\neq 0$ can be harmful or favorable for pair binding.

\begin{figure}
\includegraphics*[width=20pc]{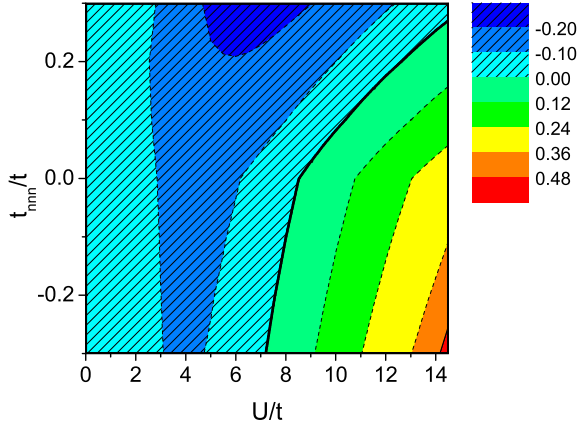}
\caption{A contour plot of charge gap $\Delta^c$ for different $U$ and $t_{nnn}$ at $T=0$. As $t_{nnn}$ increases the crossover point shifts to a larger $U$ values and the maximal charge gap width increases as $t_{nnn}>0$ but remains the same as $t_{nnn}<0$. The negative gap region is shaded.}
\label{fig:nnnT0}
\end{figure}

At zero temperature, charge gaps at one hole off half-filling (contour plots or isolines) for different $t_{nnn}$ and $U$ values are shown in Figure ~\ref{fig:nnnT0}. The charge gap behavior in the region $-0.3<t_{nnn}<0.3$ is similar to that found in section ~\ref{charge}: A negative charge gap, representing charge pairing, can be found at relatively small $U$ values. The bold line in Figure ~\ref{fig:nnnT0} shows where the charge gap becomes zero, which defines a boundary between charge pair phase and electron-hole pairing phase. As $t_{nnn}$ increases from negative (opposite sign as that of $t$) to positive (same sign as that of $t$), the boundary shifts to larger $U$ values and the magnitude of the gap also becomes larger. This can be explained by the effects of $t_{nnn}$ on the ground state with holes. The hopping between next nearest neighbors are energetically unfavorable for the half-filled antiferromagnetic state since the next nearest neighbors have the same spin direction. When holes are created in the antiferromagnetic background, hopping between next nearest neighbors will become more active and will have a measurable effect on the ground state. Depending on the sign of $t_{nnn}$, it will make the ground state with holes higher ($t_{nnn}<0$) or lower ($t_{nnn}>0$) in energy. Therefore a positive $t_{nnn}$ makes hole-hole pairing (charge pairing) ground state more energetically favorable and enhances the negative charge gap. In spite of this introduced frustration in the square system,  pairing can be enhanced
by  appropriate sign of the next-nearest-neighbor coupling ($t_{nnn}>0$).

\begin{figure}
\includegraphics*[width=20pc]{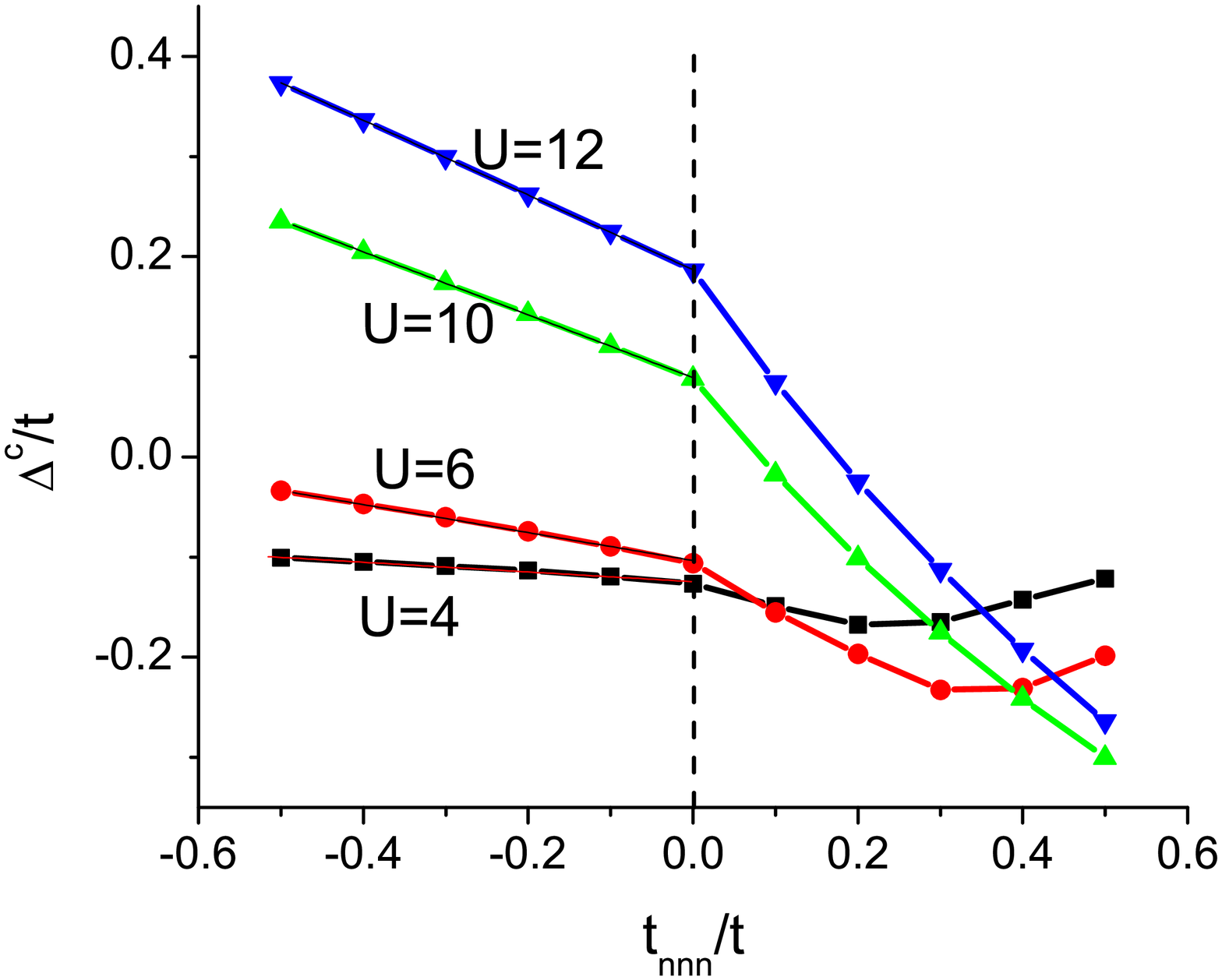}
\caption{Charge gap at fixed $U$ value for different $t_{nnn}$. A linear relation is found in the negative $t_{nnn}$ region. When $t_{nnn}>0$, the variation of the gap is not linear. This picture suggests that clusters with different signs for the next nearest neighbor hopping have different effects on the spectral properties and electron charge pairing.}
\label{fig:tpfixU}
\end{figure}

At fixed $U$, tracing charge gaps for different $t_{nnn}$ yields Figure ~\ref{fig:tpfixU}. In the $t_{nnn}<0$ region, charge gaps are linear with respect to $t_{nnn}$, while they show nonlinear behavior for $t_{nnn}>0$. The charge gap for $T=0$ is consistent with the conclusions drawn from ground state calculations in $t-t^\prime-J$ model at $t^\prime<0$ and $t^\prime>0$~\cite{White}. This linear behavior for negative $t_{nnn}$ suggests that the variation of the charge gap for small nonzero $|t_{nnn}|$ can be obtained from the case for zero $t_{nnn}$ with a linear extrapolation, which makes further research on next nearest hopping easier. Further calculations show that the linear behavior is lost when $t_{nnn}$ is positive and large; since  $t_{nnn}$ has already exceeded the range where $t_{nnn}$ is reasonably small (as the hopping between next nearest neighbors), we will not discuss it in this article. Figure ~\ref{fig:tpfixU} also shows that the linear slope approaches zero as the magnitude of the gap approaches its maximum 
(antinode) value (around $U=4$),
which indicates that, although negative $t_{nnn}$ drives charge pairing crossover point to smaller 
%U 
$U$
values, it does not affect the maximum gap
%width 
value
significantly.

\section{Summary}~\label{conclude}
We have discussed charge and spin pairing in the 8- and 10-site Betts clusters subject to periodic boundary conditions. This work provides ample evidence that such pairing instabilities do exist and are robust at the 8- and 10-site cluster sizes in the ground state and at finite temperatures. An important question is whether the obtained electron instabilities will continue to exist in the two-dimensional lattice as the cluster size increases, especially, in the thermodynamic limit. As exact calculations cannot go up to clusters large enough to eliminate size effects, the periodic Betts cells are considered to be the best optimal structures that can minimize and reduce edge effects. The key intrinsic properties of an infinite square lattice can be extracted from exact calculations in finite Betts lattices~\cite{Oitmaa}. For example, the extrapolated ground state energy from 8- and 10-site Betts cells per site at half-filling at large $U$ is quite close to existing analytical result in thermodynamical limit, $-1.15\times \frac{4t^2}{U}$~\cite{Becker}. The basic scenario of electron pairing due to the phase separation instabilities, reproduced in small Betts lattice near half-filling, can be valid also in larger size Betts lattices. Our finite size studies are also quite relevant in view of the observed magnetic and density phase separation instabilities~\cite{Fisk} and electronic inhomogeneities at the nanoscale in cuprates, pnictides, manganites and colossal magnetoresistance (CMR) nanomaterials~\cite{PLA09}.

The exact diagonalization of the Hubbard model on Betts cells (near one hole off half-filling) can be used to extract QCPs and crossover temperatures, which in turn provide important insight into the mechanisms of pairing and phase separation instabilities~\cite{AKA08P,GW09}. The exact diagonalization provides strict criteria for phase separation instabilities and boundaries between smooth and first order phase transitions~\cite{koch}. The introduction of next nearest neighbor hopping shifts the quantum crossover points but cannot eliminate the conditions necessary for electron and spin pairing. Moreover, we find that by an appropriate choice of the sign of 
%next nearest neighbor coupling (
$t_{nnn}>0$,  electron charge and opposite spin pairings in coherent phase can be strongly enhanced. The sign 
%of 
change of $t_{nnn}$ term have different effect on the spin pairing.
 %too. 
In contrast, we find a stabilization of the Nagaoka ferromagnetism with the negative spin pairing gap of the optimal value at large $U$ values in case of negative sign of $t_{nnn}<0$.

The authors  acknowledge the computing facilities provided by the Center for Functional Nanomaterials, Brookhaven National Laboratory, which is supported by the U.S. Department of Energy, Office of Basic Energy Sciences, under Contract No.DE-AC02-98CH10886. One of us (A.N.K.) thanks Edward Rezayi, Jose Rodriguez, Oscar Bernal, GuoMeng Zhao and Daniil Khomskii for useful discussions.

\end{document}